\documentstyle[12pt]{article}

\begin{document}

\begin{titlepage}
\begin{flushright}
\hfill {PSI-PR-95-06}
\end{flushright}
\begin{flushright}
\hfill {April, 1995}
\end{flushright}
\vspace{2cm}
\begin{center}

\begin{Large}
{\bf FINAL STATE INTERACTIONS OF $B\rightarrow DK$ DECAYS
}
\end{Large}

\vspace*{1cm}
	 {\bf Hanqing Zheng}\\
\vspace*{0.5cm}

	 P. Scherrer Institute, 5232 Villigen, Switzerland\footnote{e--mail:
zheng@cvax.psi.ch}

\vspace*{1.5cm}

\end{center}
\begin{abstract}

We study the final state strong interactions of the $B\rightarrow DK$
decay processes, using the
Regge model. We conclude that the final state interaction phases
are very small, typically a few degrees.
Neglecting final state interactions
in obtaining the weak decay amplitudes is a good approximation.
\end{abstract}
\vskip 1cm
\end{titlepage}

The importance of studying the final state strong interactions
of the two body non-leptonic  B decays is due to the desire of
determining the relevant CKM
matrix elements and of studying the CP violation effects
in the B decay processes. In the first case the final state strong
interactions will modify the value of the CKM
matrix elements extracted from the bare weak decay amplitudes.
In the latter case,
in order to have observable CP violation
effects, it is necessary
to find the difference between the decay rate of a process
$i\rightarrow f$ and the rate of its charge conjugate process,
$\bar i\rightarrow \bar f$ where there have to be at least two
interfering amplitudes with different weak interaction phases
 as well as
different strong interactions phases.
By writing the bare weak interaction T matrix as,
\begin{equation}
A =g_1A_1+g_2A_2  \ ,   \end{equation}
where the $g_i$ are the weak couplings, we have, after taking the
final state strong interactions into  account,
\begin{equation}
  <f|A|i>=g_1A_1e^{i\alpha_1}+g_2A_2e^{i\alpha_2}\ ,
\end{equation}
\begin{equation}
  <\bar f|A|\bar i>=g_1^*A_1e^{i\alpha_1}+g_2^*A_2e^{i\alpha_2}\ ,
\end{equation}
 where the $\alpha _i$ are the strong interaction phases. The difference
in the decay widths of
$i\rightarrow f$ and $\bar i\rightarrow \bar f$ is,
\begin{equation}
   \Gamma -\bar \Gamma \sim Im(g_1^*g_2)\sin(\alpha _1-\alpha _2)\ .
\end{equation}
  We see  that whether there are observable CP
violating effects depends crucially on the strong interaction
phase difference between the two interfering amplitudes.

Despite of the importance of the final state
strong interactions in B decay processes,
it is however difficult to deal with
and  poorly understood.
In the K decay
system the final state strong interaction
phase difference can be estimated reliably  using the low
energy effective theory of  strong interactions.
For the D decay system in which the center of mass energy is too large
to apply  the low energy effective theory,
a single resonance dominance assumption was used in studying the
final state strong interactions.
Under this assumption, the strong interaction amplitude is simply
parametrized as the function of the decay rate of the resonance
and the couplings between the resonance and the interacting particles.
The phase difference can therefore be estimated in a resonable range
of  parameter space~\cite{D}.
This method can not  be successful
for  the B decays because we know from QCD that
the number density of resonances is an increasing function
of the mass.

In this paper, we study the final state
strong interactions of the two body non--leptonic B decay processes
using the Regge model
analysis~\footnote{A brief discussion based on
the Regge model was given in ref. \cite{Dib}}.
This method is based upon the
strong interaction duality argument: When the center of mass energy
gets high enough, the summation  of the contributions from  the
s--channel resonances to the amplitude
is equivalent to the summation of the contributions
from leading  t--channel Reggeon exchanges.
The Regge model has been proven to be very successful
in explaining high energy hadron scattering in the small $-|t|$
region~\cite{collins}.
As an example, in the case of $\pi N$  elastic scattering
resonance models only work when the center of mass energy
is less than $s\simeq 6GeV^2$ and fail badly when $s$ exceeds $10GeV^2$
where the Regge model starts to work very well ~\cite{piN} (In charge
exchange processes it works even at lower energies).
In order to have a clear insight on the physics  concerned,
we  limit ourselves to the processes without penguin diagram
contributions. The absorbtion part of the amplitudes in such a case come
 only from "soft" final state
strong interactions. Especially, we discuss
the $B\rightarrow D\bar K$ and $B\rightarrow \bar D\bar K$ processes
although  applications of our method to any two body decay process are
straightforward.

The relation between the full physical amplitude when the
final state interactions are taken into account and the bare
amplitude is given by the final state theorem of Migdal and
Watson~\cite{watson},
\begin{equation}\label{3.1}
A_{tot}=S^{1\over 2}A_{bare}\ .
\end{equation}
where $S$ is the final state strong interaction S matrix for the
given partial wave ($J=0$ in the present case).

The final states are classified according to
their flavor quantum numbers.
For a fixed final state f various intermediate states can contribute
coherently to the rescattering.
For example for the final state $D^+K^-$
both $D^+K^-$ and $ D^0\bar K^0$ are possible as intermediate states
leading to a $2\times 2$ $S$ matrix.
While for $\bar D^0
K^-$, $D^-\bar K^0$ case things are more complicated since the
$D_s^-\eta$ and $D_s^-\pi^0$ also contribute and
the $S$ matrix is  $4\times 4$ \footnote{There are many other
states with the same total
spin and flavor quantum numbers, including two body
states with excited particles and multi-particle states.
As an approximation we systematically neglect them. We will discuss their
influence later in this paper.}.
The Pomeron and Reggeon contributions to the s channel
scattering amplitude
A(s,t) can be written down systematically using the crossing
symmetry between the s channel and the t channel amplitudes,
the line-reversal law~\cite{collins},
and the $SU(3)$ relations~\cite{Rosner},
\begin{eqnarray}
\label{3.00}
A(\bar D^0K^-\rightarrow \bar D^0 K^-)=
& P + \rho + A_2 +\omega + f \ ,& \\
A(\bar D^0\bar K^0\rightarrow \bar D^0 \bar K^0)=
& P - \rho - A_2 +\omega + f \ ,& \\
A(D^- \bar K^0\rightarrow D^- \bar K^0)=
& P + \rho + A_2 +\omega + f \ ,& \\
A(D^+ \bar K^0\rightarrow D^+ \bar K^0)=
& P - \rho + A_2 -\omega + f \ ,& \\
A(D^+      K^-\rightarrow D^+      K^-)=
& P + \rho - A_2 -\omega + f \ ,& \\
A(     D^0\bar K^0\rightarrow      D^0 \bar K^0)=
& P + \rho - A_2 -\omega + f \ ,& \\
A(     D^0     K^-\rightarrow      D^0      K^-)=
& P - \rho + A_2 -\omega + f \ ,& \\
A(     D_s^-   \pi^0\rightarrow      D_s^-    \pi^0)=
& P \ ,& \\
A( D_s^- \pi^+\rightarrow D_s^-\pi^+)=&P\ ,&\\
A(     D_s^-   \eta\rightarrow      D_s^-    \eta)=
& P+{8\over 3}f'\ ,&
\end{eqnarray}
where the P, $\rho$, $\omega$, $A_2$,
$f$ and $f'$ denote the contribution
from the Pomeron, $\rho$, $\omega$, $A_2$, $f$ and $f'$
 Reggeon exchanges to
the the amplitude, respectively.
For the charge exchange processes, we have
\begin{eqnarray}
\label{3.01}
A(\bar D^0K^-\rightarrow D^- \bar K^0 )= & 2\rho + 2A_2 \ ,& \\
A(D^+  K^-\rightarrow D^0 \bar K^0)= &  -2\rho + 2A_2  \ ,& \\
A(D_s^-\eta\rightarrow \bar D^0 K^- )= &-\sqrt{6}K^*
-\sqrt {2\over 3}K^{**}\ ,& \\
A(D_s^-\eta\rightarrow D^-\bar K^0 )= &-\sqrt{6}K^*-
\sqrt {2\over 3}K^{**}\ ,& \\
A(D_s^-  \pi^0\rightarrow D^- \bar K^0)
= &\sqrt {2}(K^* - K^{**}) \ ,& \\
A(D_s^-  \pi^0\rightarrow \bar D^0 K^-)
= &\sqrt{2}(K^{**} - K^{*}) \ ,& \\
A(D_s^-\pi^+\rightarrow \bar D_0\bar K_0)=&2(K^{**}-K^*)\ ,&
\end{eqnarray}
where the $K^*$ and $K^{**}$ denote the $K^*$ and $K^{**}$ Reggeon
contributions, respectively. The $SU(3)$ coefficients are written
explicitly in above formulas.
Please notice that the Pomeron couplings
may depend on the different flavour content  while
various Reggeon contributions are related to each other by
the strong exchange degeneracy (SED) requirement~\cite{collins}.
Especially, the
magnitude of each Reggeon contribution in an amplitude
 should be equal.
 The SED
 is known to
 hold quite  well  for vector and tensor
 Reggeon exchange processes  especially
 when the strange quark is involved.
 Sign differences between Reggeon
 contributions in  the above formulas
lead to the cancellation between the
imaginary part of the Reggeon contributions to  s channel exotic
amplitudes.
 This is just a manifestation of the  fact
 that  the absence
 of an imaginary part in $A^R$
 implies that there is no s channel resonance,
because of the duality argument.

 We parametrize the Pomeron and the Reggeon exchange amplitudes
 in the small $-|t|$ region in the following way:
\begin{equation}\label{3}
     P=\beta^P (t) \left( {s\over s_0}\right) ^{\alpha ^P(t)}
     e^{-i{\pi\over 2}
\alpha ^P(t)}\ ,\end{equation}
\begin{equation}\label{3.2}
R=\beta^R(t) \left({s\over s_0}\right)
^{\alpha ^R (t)}
{\pm 1- e^{-i\pi\alpha ^R(t)}\over \sin \pi \alpha _0^R}\ ,\end{equation}
\begin{equation}\label{3.2*}
\beta^P(t)=\beta^Pe^{a^Pt}\ \,\, \hbox{and} \,\,\,
 \beta^R(t)= \beta^R e^{a^Rt}\ , \end{equation}
 where $\pm $ sign refers
  to odd/even signatures of the exchanging Reggeons.
 The $SU(3)$ invariant couplings $\beta ^R$ are normalized such that it
is just the value of the residue function of the
Reggeon--matter coupling  at origin, i.e., $t=0$. In accordance with
the traditional treatment we use an exponential form of parametrization
to characterize the form factor of the Reggeon--matter couplings.
The parameter $a^R$ can be estimated using the Veneziano model
\cite{Veneziano} for meson-meson scatterings,
\begin{equation}\label{3.3}
a^R=-{\Gamma '(1-\alpha ^0_R)\over \Gamma (1-\alpha ^0_R)}\alpha '_R
\end{equation}
which obeys approximately the  relation:
$a^R= {1\over m_R^2}$  where $m_R$ is the mass of the corresponding
elementary particle of the Reggeon R. We will approximate
$a^R$ by $1/m_R^2$ in the following.

It is worth pointing out that our parametrization,
eq.~(\ref{3.2}), is different from what is usually used in the
literature:
\begin{equation}\label{3.3*}
A^R(s,t)=-x\beta ^Re^{a^Rt}\left( {s\over s_0}\right)
^{\alpha ^R(t)}e^{-i\pi \alpha ^R(t)/2}
\end{equation}
\noindent with $x=1/-i$ for even/odd signatures, respectively.
The reason is that as $\alpha ^0_R\ne 0.5$, the parametrization
eq.(\ref{3.3*}) violates SED
by introducing  an
imaginary part to the Regge amplitude  for a s--channel
exotic process and is therefore not adequate for the present discussion.
For the trajectory  functions $\alpha ^P(t)$ and $\alpha ^R(t)$
we use the following values \cite{collins,landshoff2},
\begin{equation}\label{3.4}
\alpha ^P(t)\simeq 1.08 +0.25t\ ,\end{equation}
\begin{equation}\label{3.5}
\alpha ^\rho (t)\simeq 0.44 +0.93t\ ,\end{equation}
\begin{equation}\label{3.6}
\alpha ^{K^*} (t)\simeq  0.3 +0.93t\ ,\end{equation}
\begin{equation}\label{3.7}
\alpha^{f'}(t)= 0.1 + 0.93t\ .
\end{equation}
To estimate   the residue functions
$\beta ^P(t)$ and $\beta^R(t)$ we assume
factorization. That is, for a process $AB\to CD$ we have,
\begin{equation}\label{3.9}
\beta _{AB\to CD}(t)=\beta _{AC}(t) \beta _{BD}(t)\ ,
\end{equation}
\noindent both for $\beta ^P(t)$ and $\beta^R(t)$.
In the literature, the Pomeron coupling to the pion is taken as
a single pole form in the small $-|t|$ region~\cite{landshoff1},
\begin{equation}\label{3.10}
\beta ^P_\pi(t)=\beta ^P_\pi {1\over 1-t/0.71}\ .
\end{equation}
The t dependence of
Pomeron couplings to $D$ and $K$ mesons are not known experimentally,
we simply
assume they are the same as that in eq.(\ref{3.10})\footnote{Roughly
speaking, the formfactor is the Fourier transformed version of the
size of the hadron. Since the $\pi$, $D$ and $K$ mesons
both involve light quarks their size is expected to be similar. }.
For the residue function of the
Pomeron amplitude $\beta^P(t)$, we take
\begin{equation}\label{3.9*}
\beta^P(t)=\beta^P_D\beta^P_K\simeq\beta^P_\pi\beta^P_\pi
\sim    \left( {1\over 1-{t/ 0.71}}\right) ^2\simeq
e^{a^Pt}=e^{2.82t}\ .
\end{equation}
The value of   $\beta ^P_{DK}$
 can be estimated as follows.
{}From $\pi p$ and $Kp$ data one gets,
\begin{equation}
\beta^P (su)\simeq {2\over 3}\beta^P (uu)\ ,
\end{equation}
and
\begin{equation}
\beta^P (cu)\simeq {1\over 10}\beta^P (uu)\ ,
\end{equation}
from \cite{landshoff2}.
Using isospin invariance
\begin{equation}
\beta ^P(uu)=\beta ^P(ud)=\beta ^P(dd)\ ,
\end{equation}
and the additive quark counting rule
we obtain
\begin{equation}
\beta^P(DK)=\beta^P(cu)+\beta^P(cs)+\beta^P(us)+\beta ^P(uu)
  \simeq 12.0 \ .
\end{equation}

The coupling constant
$\beta ^R$ in eq.~(\ref{3.2*}) cannot be estimated from
$SU(3)$ constraints, since the $D$ meson is a $SU(3)$ triplet and may
have different coupling comparing with the $SU(3)$ octet mesons.
It  can however be fixed from the $SU(4)$ relation
\begin{equation}\label{3.11}
\beta ^R_D=\beta ^R_K\ ,
\end{equation}
provided that the coupling
constant $\beta^R_K$  can be estimated from $NN$, $\pi N$ and $KN$
scattering data.  Eq.(\ref{3.11}) is
also a
direct consequence  of the vector meson coupling universality, which
is respected satisfactorily  in the $\pi (K) N$ processes.

The Regge pole amplitudes
for $\pi (K) N$ elastic scattering can be written down
analogously.
We have fitted the experimental data on the $\pi (K) N$
total cross--sections and find that $\beta^R\simeq 3.11$.

At this stage we are able to calculate the two body scattering
Regge amplitudes given above
which is however not directly
applicable for  studying  the
final state interactions of the B decay system.
Because
the $B$ meson has spin 0, the corresponding scattering
amplitude should be the s--partial wave projection of the full Regge
amplitude:
\begin{equation}\label{s1}
 A_{J=0}(s)={s\over 16\pi \lambda }\int^0_{-t_-}A(s,t)dt\ ,
\end{equation}
\noindent where
\begin{equation}
\lambda \equiv \lambda (s,m_1^2, m_2^2)= s^2 + m_1^4 + m_2^4
-2sm_1^2 -2sm_2^2 -2m_1^2m_2^2\ ,
\end{equation}
\noindent and
\begin{equation}
t_{-}={\lambda \over s}\ .
\end{equation}

Before performing the numerical calculation we first remark
on the reliability
 of our method. It is known that
the Regge parameterization is only
valid in the small $-|t|$
 region. In order to obtain the s--wave amplitude
we need to perform the integration over the whole -$|t|$ region,
and further,
low partial wave projections of the Regge amplitudes
are considered as the most unreliable part of the theory.
 However,
the large $-|t|$ contributions to the
total scattering amplitude   is suppressed by 1/s,
therefore the uncertainty due to the
 invalidity of the Regge amplitude at large $-|t|$ region
and the dependence of our results on the parameterization form
in the small $-|t|$ region should not be important.
For the large $-|t|$ region
 Regge cuts and exchange channel Reggeons dominate.
The Regge trajectory in the u--channel
 are $D^*$ and $D^{**}(2460)$ in our case.
However, these u--channel
 Reggeons
have a very small intercept $\alpha ^0_R$:
\begin{equation}\label{d1}
\alpha ^0_{D^*}=\alpha ^0_{D^{**}}=
1-{m^2_{D^*}\over  m^2_{D^{**}}-m^2_{D^*}}\simeq -1 \ ,
\end{equation}
\noindent their contributions are therefore negligible.

For the  Regge cuts, it was realized long ago that
one  of the
main defects of Regge poles
is that they give too large a contribution
to the low partial waves. Some extra absorption  provided
by the Regge cuts is necessary. However, the absorption effects in our
case (meson--meson scattering) should not be large either.
There is strong experimental evidence indicating that the absorption
effects are much smaller in  $\pi N$ scattering  than in
$NN $ scattering. It is natural to expect that in meson meson
scattering  these absorption effects are even smaller.
This may be  understood from  the simple
Reggeized absorption model~\cite{Kane}
despite of the fact that Regge cuts  are less well
known theoretically than Regge poles. In
the simple Reggeized absorption model  the cuts
generated by two Pomerons or one Pomeron one Reggeon exchange contribute
to the full s--partial wave amplitude as:
\begin{equation}\label{d2}
A^{P+P\bigotimes P}=A^P\left( 1- {\lambda_P \beta ^P\over 16\pi c_Ps_0}
\left( {s\over s_0}\right)^\epsilon
\right)\ ,
\end{equation}
and
\begin{equation}\label{d3}
A^{R+R\bigotimes P}=A^R\left( 1- {\lambda_R \beta ^P\over 8\pi c_Ps_0}
\left( {s\over s_0}\right)^\epsilon e^{-i{\pi\over 2}\epsilon}
\right)\ ,
\end{equation}
\noindent where $\epsilon =0.08$,
$c_P= a^P +\alpha _P'\left( \log ({s\over s_0}) -{\pi/2}i\right)$.
The phenomenological enhancement factor $\lambda$  characterizes
the contribution from quasi--elastic intermediate states.
We read off from eqs.(\ref{d2}), (\ref{d3}) that the  contribution
from the cuts
is proportional to $\beta ^P$ which is much smaller in our case,
$\beta ^P_{DK}\simeq 12$, than in the case of $\pi N$ scattering.
This is crucial to reduce significantly the correction to the s
wave amplitude, different from the $\pi N$ scattering case.
For example,
taking $\lambda _R =2$ in the latter  case will lead to
a complete absorption in the s--wave amplitude, $A_s\sim 0$ while the
same value of the $\lambda_R$
parameter only reduces $A_s(DK)$ by $\sim
30\%$. Even though we have poor knowledge in estimating
the $\lambda_R$
parameter we expect this amount of reduction
is resonable,
as an upper limit.
The $\lambda _P$ parameter may be safely neglected.
According to
ref.~\cite{landshoff1} the inclusion of the P$\bigotimes$P cut
in the  $NN$ scattering case
 only leads to a small change
of the $\beta$ parameter.

Because of the above arguments we use in the following only  the
simple Regge pole model
to evaluate the effects of the final state interactions of the B decay
processes from the  final state theorem, eq.(\ref{3.1}).
In the  practical calculation it is convenient to study the problem
in the strong interaction eigenstates
for which  the $S$ matrix has a diagonal form. For the $D \bar K$
case we have
  \begin{equation}
  A(D\bar K)=\left( \matrix{A_{I=0}(D\bar K)&0
   \cr 0& A_{I=1}(D\bar K)\cr}\right)\ ,
\end{equation}
where
$A_{I=0}= P+2(\rho -A_2)$, $A_{I=1}= P+2(A_2-\rho ) $ and
$|D\bar K>_{I=0}={1\over \sqrt{2}}(|D^+K^->-|D^0\bar K^0>$,
$|D\bar K>_{I=1}={1\over \sqrt{2}}(|D^+K^->+|D^0\bar K^0>$.
For the $\bar D\bar K$ case,
there are two degenerate states of each isospin
which are,
\begin{equation}
|0^1>= -{1\over \sqrt {2}}(|\bar D^0 K^-> + |D^-\bar K^0>),\
  |0^2>=|D_s^-\eta >,\end{equation}
\begin{equation}
|1^1>=-{1\over \sqrt {2}}(|\bar D^0 K^-> - |D^-\bar K^0>),\
  |1^2>=|D_s^-\pi^0>.\end{equation}
The $J=0$ $S$ matrices,
$S_{ij} =\delta_{ij}+2i\sqrt{\rho_i\rho_j} A^{J=0}_{ij} $
 ($\rho_i =\sqrt{\lambda_i}/s$) are the following:
\begin{equation}
\left(\matrix{<0^1|0^1>&<0^1|0^2>\cr <0^2|0^1> & <0^2|0^2>\cr}\right),\
\left(\matrix{<1^1|1^1>&<1^1|1^2>\cr <1^2|1^1> & <1^2|1^2>\cr}\right).
\end{equation}
In the  strong interaction eigenstates $S$
is parametrized as,
\begin{equation}
S^{J=0}\equiv diag(\eta_{I}e^{2i\delta _I}),\end{equation}
The $\eta_I$ parameter characterizes the inelasticity
of the given process.

The numerical results for the parameters given  above
are\footnote{The $S$ matrix is equal to,
$$\left(\matrix{a&b\cr b & a'\cr}\right)\ ,$$
with $a-a', b\sim  O(10^{-1})a$. The $S^{1/2}$ matrix is of the
following form:
$$\left(\matrix{x&y\cr y & x'\cr}\right)\ ,$$
with $x,x'>>y$. Approximately  $x=\sqrt{a}$, $y=b/2\sqrt{a}$ and
$x'=\sqrt{a}+ (a'-a)/2\sqrt{a}$. The numerical results are,
$$ S^{1/2}_{I=0}(\bar D\bar K)\simeq
\left(\matrix{0.87-3.58i\cdot 10^{-2}&
1.11\cdot 10^{-2}-4.19i\cdot 10^{-2}\cr
1.11\cdot 10^{-2}-4.19i\cdot 10^{-2}
& 0.95-4.40i\cdot 10^{-2}
\cr}\right)\ ,$$
$$
S^{1/2}_{I=1}(\bar D\bar K)\simeq
\left(\matrix{
0.91+3.06i\cdot 10^{-3}& 2.76i\cdot 10^{-2}\cr
2.76i\cdot 10^{-2}     & 0.93 +2.52i\cdot 10^{-3}
\cr}\right)\ .$$
We see that the nondiagonal elements are much smaller than the diagonal one.},
\begin{equation}\label{data1}
\eta _0(\bar D\bar K)=0.81,\ \delta _0=-3.3^\circ ;
\eta _1(\bar D\bar K)=0.84, \delta _1=1.8^{\circ}; \\
\end{equation}
\begin{equation}
\eta _0( D_s^-\eta)=0.85,\ \delta_0 =-1.7^\circ ;
\eta _1( D_s^-\pi^0)=0.83, \delta _1=-1.5^{\circ};
\end{equation}
\begin{equation}\label{data2}
\eta _0(D\bar K)=0.83,\ \delta _0=2.2^\circ ;
\eta _1(D\bar K)=0.83\ , \delta _1=-1.8^\circ .
\end{equation}

{}From above we see that
the phase shifts $\delta $ are very small (modulo $n\pi$, of course).
This can be understood
qualitatively: At the center of mass energy $m_B$
the Pomeron contribution is dominant which gives
almost a purely imaginary
contribution to the $A$ amplitude and therefore,
the $S$ matrix elements are
almost purely real. Furthermore, in our case
the Pomeron contribution is much smaller than in the $\pi N$ or $NN$
case because of the smaller couplings.
Only small cancellation occur in the $S$ matrix elements between
the Pomeron contribution (mainly a negative real value) and
unity (the value of the $S$ matrix element in
the limit of vanishing interaction).

We notice that in our results the $\eta$ parameters
are slightly less
than 1, which indicates
that we have neglected some final states in addition to
 those  considered. In order to restore unitarity
one has to take them into
account which will  lead to corrections to the relations between the
bare amplitudes and the full amplitudes  obtained above. The
calculation of the effects brought by  these states
go beyond the ability of the present model analysis.
In the following we try to give a simple qualitative estimate.

Consider a given strong interaction eigenstate $|a>$
which diagonalizes the $S$ matrix in the incomplete  basis like
the states that we have considered: $<a|S|a>=\eta e^{2i\delta_1}$
and $\eta <1$.
Since $\eta$ is less than 1  unitarity  tells us that in
the complete basis the $S$ matrix is still nondiagonal. Many new states
can contribute to restore unitarity by introducing the non-diagonal matrix
element $<new|S|a>$.
Now we assume that all the effects of these states are characterized  by a
single effective state $|a'>$ which, together with $|a>$ leads to a
$2\times 2$ unitary
$S$ matrix which can be simply parametrized as,
\begin{equation}
\left(\matrix{\eta e^{2i\delta_1} &
i\sqrt{1-\eta^2}e^{i(\delta_1+\delta_2)
} \cr i\sqrt{1-\eta^2}e^{i(\delta_1+\delta_2)}
& \eta e^{2i\delta_2}\cr}\right)\ .
\end{equation}
The solution of $S^{1/2}$ is
\begin{equation}
\left(\matrix{x e^{i\alpha_1} &  i\sqrt{1-x^2}e^{i
{(\delta_1+\delta_2)\over 2}
} \cr i\sqrt{1-x^2}e^{i{(\delta_1+\delta_2)\over 2}}
& xe^{i\alpha_2}\cr} \right) \ .
\end{equation}
where
\begin{equation}
\label{x}
x=\sqrt{{1+2\eta\cos(\delta_1-\delta_2)+\eta^2\over
2(1+\eta\cos(\delta_1-\delta_2))}}
\end{equation}
and
\begin{equation}
\label{alpha}
\alpha_{1,2}={\delta_1+\delta_2\over 2} \pm \sin^{-1}\left(
{\eta\sin(\delta_1-\delta_2)\over
\sqrt{1+2\eta\cos(\delta_1-\delta_2)+\eta^2}}\right)\ .
\end{equation}
The full physical amplitude we are interested in $<a|B>^f$ can then be
written as,
\begin{equation}
\label{fb}
<a|B>^f=x e^{i\alpha} <a|B>^b +\sqrt{1-x^2}
ie^{ i({\delta_1+\delta_2\over 2}) }<a'|B>^b\ .
\end{equation}

\noindent From eqs.~(\ref{x}) and (\ref{alpha})
we see that x is very close to 1 and $\alpha_1$
very close to $\delta_1$ when
$\eta\sim 0.8$ within a reasonable range of the $
\delta_2 $ parameter, say,
$|\delta_2|<30^\circ$
(remember that $\delta_2 $ should also not be very
large because of the Pomeron
dominance). Therefore the phase shift
$\alpha_1$  remains small. This is in agreement with the
present experimental evidence. Furthermore
 since $<a'|B>$  represents an averaged
value of contributions from many states it is expected that  large
cancelation between different amplitudes should occur which leads to
$|<a'|B>/<a|B>|<<1$. Therefore we can conclude from
eq.~(\ref{fb}) that the bare weak decay amplitude is a good
approximation to the  full physical amplitude, accurate
up to, roughly speaking, about 10 percent.

To conclude, the partial wave elastic unitarity is a good approximation
for the $DK$ scatterings at $s=M_B^2$ even though we know that "Pomeron
dominance" implies inelasticity. In our special
case it happens that these two statements are consistent.
One simple way to understand  qualitatively   the
smallness of $\Delta \delta_I$ ($=\delta_1-\delta_0$) is that
only the Reggeon contribution is isospin dependent. Therefore it is
proportional to $\Delta R_I/(1-P)$ and be a small quantity.
The above analysis can be applied to other processes. In $D\pi$ processes
similar qualitative results should hold.
In the $\pi\pi$ and $K\pi$
cases the Pomeron contribution gets
 larger (by a factor of $\sim 2$),
the value of
the $\eta$ parameter is further decreased ($\eta \sim 0.6 - 0.7$).
Uncertainties
raised from  inelasticity are more serious
and sound conclusions are difficult to make.

Acknowledgment:
I have much benefited from conversations with M. Jacob,
P. Landshoff, G. Levin, M. Locher, J. Rosner and G. Veneziano.
Special thanks are given to I. Dunietz, who  has taken part in the early
stage of this work, for his  valuable contributions.

\end{document}